# A Method to determine Partial Weight Enumerator for Linear Block Codes

Saïd NOUH[1*], Bouchaib AYLAJ[2] and Mostafa BELKASMI[1]

1. Mohammed V-Souisi University, SIME Labo, ENSIAS, Rabat, Morocco.
2. Chouaib Doukkali University, MAPI Labo, Faculty of Science, El jadida, Morocco

* E-mail of the corresponding author: nouh_ensias@yahoo.fr

**Abstract**

In this paper we present a fast and efficient method to find partial weight enumerator (PWE) for binary linear block codes by using the error impulse technique and Monte Carlo method. This PWE can be used to compute an upper bound of the error probability for the soft decision maximum likelihood decoder (MLD). As application of this method we give partial weight enumerators and analytical performances of the BCH(130,66) , BCH(103,47) and BCH(111,55) shortened codes; the first code is obtained by shortening the binary primitive BCH (255,191,17) code and the two other codes are obtained by shortening the binary primitive BCH(127,71,19) code. The weight distributions of these three codes are unknown at our knowledge.

**1. Introduction**

The current large development and deployment of wireless and digital communication has a great effect on the research activities in the domain of error correcting codes. Codes are used to improve the reliability of data transmitted over communication channels, such as a telephone line, microwave link, or optical fiber, where the signals are often corrupted by noise. Coding techniques create code words by adding redundant information to the user information vectors. Decoding algorithms try to find the most likely transmitted codeword related to the received word as illustrated in the figure 1.

Decoding algorithms are classified into two categories: Hard decision and soft decision algorithms. Hard decision algorithms work on a binary form of the received information. In contrast, soft decision algorithms work directly on the received symbols (Clark 1981).

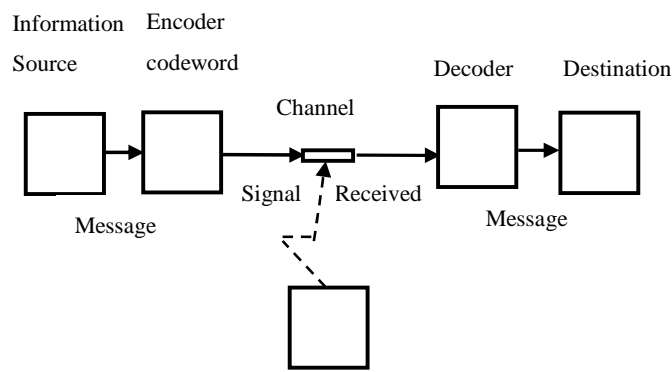

Figure1. A simplified model a communication system.

Let C(n,k,d) be a binary linear block code of length n, dimension k and minimum distance d. The weight enumerator of C is the polynomial:

$$A(x) = \sum_{i=0}^{n} A_i x^i$$

Where $A_i$ denotes the number of codewords of weight i in C. The weight of a word is the number of non-zero elements it contains. The polynomial A is one of the keys to obtain an exact expression for the error detection and error correction performance of C (Clark 1981 and Berlekamp 1984). It can be used to compute an upper bound on the error correcting performances for the soft decision maximum likelihood decoding algorithm (MLD): The probability of a decoding error with the MLD decoder denoted by $p_e$ can be upper bounded as follows (Clark 1981):

$$p_e(C) \leq \sum_{w=d}^{n} A_w Q\left(\sqrt{2wR\frac{E_b}{N_0}}\right) \quad (1)$$

Where A is the weight enumerator of C, $R=k/n$ is the code rate, $E_b/N_0$ is the energy-per-bit to noise ratio and $Q(x)$ is given by:

$$Q(x) = \frac{1}{\sqrt{2\pi}} \int_x^{\infty} e^{-z^2/2} dz \quad (2)$$

The error correcting performances obtained by Equation (1) are called the analytical performance of C.

In (Fossorier 1998), authors show that for systematic binary linear codes with binary transmission over AWGN channel, the probability of bit error denoted by $P_b(C)$ or BER(bit error rate), has the following upper bound:

$$p_b \leq \sum_{w=d}^{n} \frac{wA_w}{n} Q(\sqrt{2wR\frac{E_b}{N_0}}) \quad (3)$$

The above bound holds only for systematic encoding; this shows that systematic encoding is not only desirable, but actually optimal in the above sense.

In general the determination of the weight enumerator is a NP-hard problem; the simplest way is the exhaustive research which requires the computing of $2^k$ codewords.

For some codes, the wealth of their algebraic structures permits to determine the weights enumerators; so many excellent studies on the determination of the weights enumerators of quadratic codes or their extended codes were succeed (Gaborit 2005, Tjhai 2008, Su 2008, Mykkeltveit 1972) and the weights enumerators are known for all lengths less than or equal to 167.

In (Fujiwara 1993) authors have found the weight enumerators of the primitive BCH codes of length 255 and dimension less than 64 or big than 206.

In (Sidel'nikov 1971) Sidel'nikov has proved that the weight enumerator of some binary primitive BCH codes can be given by the approximate following formula:

$$A_j = 2^{-mt} \binom{n}{j} \cdot (1+R_j)$$

Where $|R_j| \leq K \cdot n^{-0.1}$ and K is a constant. n is the code length in the form $2^m-1$ and t is the error correction capability of the code. In (Kasami 1985) Kasami et al have given an approximation of the weight enumerator of linear codes with the binomial distribution by using a generalization of the Sidel'nikov result.

The BCH codes, as a class, are one of the most known powerful error-correcting cyclic codes. The most common BCH codes are characterized as follows: specifically, for any positive integer $m \geq 3$, and $t < 2^{m-1}$, there exists a binary BCH code with the following parameters:

- Block length : $n=2^m-1$
- Number of message bits : $k \leq n-mt$
- Minimum distance : $d \geq 2t + 1$

These BCH codes are called primitive because they are built using a primitive element of GF($2^m$).

The code obtained by removing a fixed coordinates of a primitive BCH code is called a shortened BCH code. Thus the BCH(130,66) code is obtained by removing 125 coordinates of the binary primitive BCH(255,191,17) code; the BCH(103,47) and BCH(111,55) codes are obtained by removing respectively 24 and 16 coordinates of the binary primitive BCH(127,71,19) code.
In this paper an approximation of partial weight enumerators of the BCH(130,66), BCH(111,55) and BCH(103,47) codes are given.

In (Nouh 2011), authors have used genetic algorithms combined with the Monte Carlo method for finding the weight enumerator of the quadratic residue code of lengths 191 and 199. This method is efficient when the linear system obtained from the Mac Williams identity (MacWilliams 1977) contains a small number of unknowns after its simplification (Nouh 2011). For the BCH(130,66) code it remains **37** unknowns then the method given in (Nouh 2011) can't be used. So we introduce here a new method allowing determination of a partial weight distribution of a block code.

The remainder of this paper is organized as follows. On the next section, we define the partial weight enumerators of a code and the corresponding analytical performances. In section 3 we present the proposed method for approximating a partial weight enumerator. In section 4 we give a validation of the proposed method and we give their results for some BCH codes. Finally, a conclusion and a possible future direction of this research are outlined in section 4.

*2.* **Partial weight enumerator and analytical performances.**

2.1 *Union bound on the error probability of the Maximum likelihood Decoding algorithm (MLD).*

Let be p a positive integer less than n-d. The polynomial $A^p(x) = 1 + \sum_{i=d}^{p+d} A_i X^i$ is called the partial weight enumerator of radius p of the code C having the weight enumerator A.

Let be the function $P_e(p)$ defined by Equation (4):

$$P_e(p) = \sum_{w=d}^{p} \frac{w A_w}{n} Q(\sqrt{2wR\frac{E_b}{N_0}}) \qquad (4)$$

The Equation (3) becomes:  $\qquad p_e \leq p_e(p) + \varepsilon \qquad (5)$

The Equation (2) shows that $p_e(p)$ increase slowly with p and there exists an appropriate value of p for which the value of $\varepsilon$ becomes negligible comparing to $p_e(p)$ therefore the code C performances can be analyzed by using a partial weight enumerator of radius p. In this paper we will show by simulation that the appropriate value of p can be chosen between 2 and 4 without big alteration of the analytical performances.

We have firstly decoded the extended (24,12,8) Golay code having the weight enumerator:

$A(X) = 1 + 759X^8 + 2576X^{12} + 759X^{16} + X^{24}$ by the MLD decoder. The figure 2 shows a comparison between the simulated error correcting performance of the MLD decoder and the analytical performance AP(WE) obtained by Equation (3) and the weight enumerator (WE=A). Practically the performances are the same for the values of SNR greater than 2. The simulation parameters of the MLD decoder are given in the table 1.

Table 1. The simulation parameters.

| Simulation parameter | value |
| --- | --- |
| Channel | AWGN |
| Modulation | BPSK |
| Minimum number of residual bit in errors | 200 |
| Minimum number of transmitted blocks | 5000 |

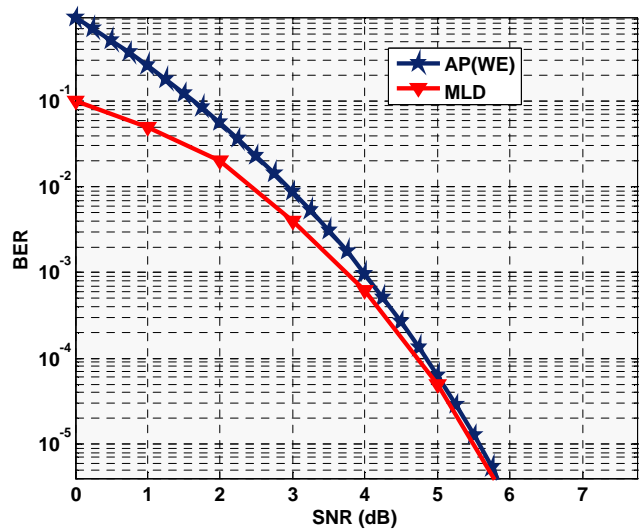

Figure 2. Comparison between simulated and analytical performances of the MLD decoder for the extended (24,12,8) Golay code.

The figure 3 shows a comparison between the analytical performances AP (WE) obtained by Equation (3) when the weight enumerator (WE=A) is completely used and the analytical performances AP (PWE) obtained by the same formula with only the first two terms of A (PWE=$1+759X^8$). Practically the obtained analytical performance is the same for all values of SNR.

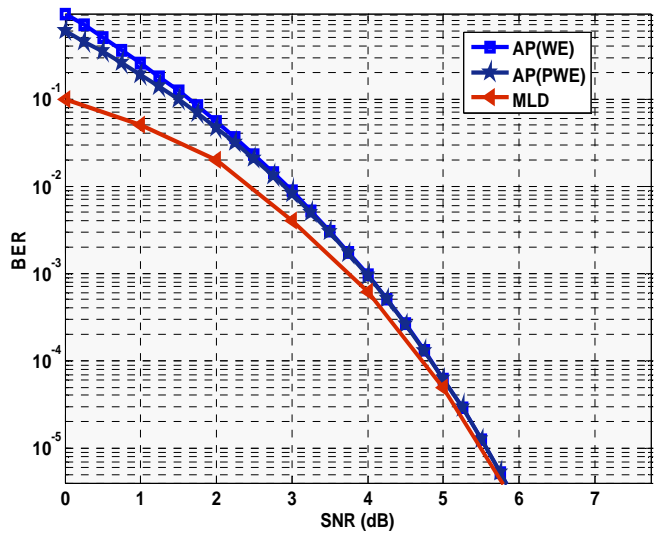

Figure 3. Comparison of the analytical performances obtained by WE and PWE for the extended (24,12,8) Golay code.

To compare the analytical performances obtained by the weight enumerator and those obtained by a partial weight enumerator of radius p between 4 and 8 for relatively long codes we give in the figure 4 the analytical performances for the BCH(127,57,23) code by using only the first p terms of its weight enumerator given in (Desaki 1997). Practically the performances are the same for all values of SNR greater than 3 dB.

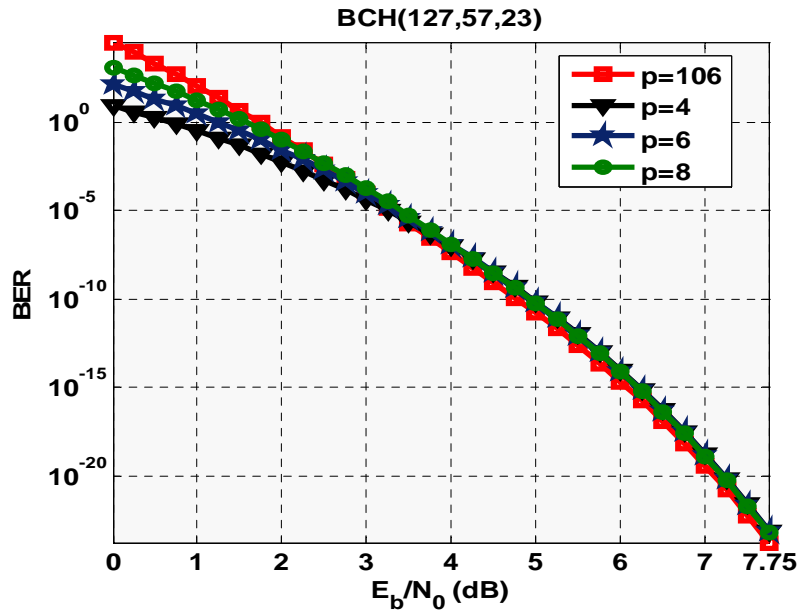

Figure 4. Impact of the radius p on the analytical performances of the BCH (127, 57, 23) code.

**3. The proposed method for finding a partial weight enumerator.**

*3.1 Monte Carlo method.*

In the general case, the Monte Carlo method consists in formulating a game of chance or a stochastic process which produces a random variable whose expected value is the solution of a certain problem. An approximation to the expected value is obtained by means of sampling from the resulting distribution. This method is used for example in the evaluation of some integrals (Bauer 1958).

*3.2 The proposed method for approximating a partial weight enumerator of linear block codes.*

In order to have a good approximation of the partial weight enumerator, we propose the following probabilistic method, which is a variant of Monte Carlo method.

Let $C_w$ be the set of all codewords of weight w in C and $L_w$ a subset of $C_w$. We define the recovery rate of $L_w$ related to $C_w$ by:
$$R(L_w) = |L_w|/|C_w| \qquad (6)$$

Where the symbol $|.|$ denotes the cardinal.

When $L_w$ is available, the value of $R(L_w)$ can be computed statistically with a percentage of error $\mu$ and $|C_w|$ can be approximated by $|L_w|/R(L_w)$.

When a codeword c of weight w is drawn at random, there are two cases:

- c is an element of $L_w$ and it is called intern code word.
- c isn't an element of $L_w$ and it is called extern code word.

13

In the practice the recovery rate $R=R(L_w)$ can be evaluated statistically by the following algorithm:

---

**Inputs**: $L_w$, M (minimum number of intern code words)

**Outputs**: R (recovery rate)

*Begin*

$S \leftarrow 0;\ i \leftarrow 0;$

*While (S < M) do:*

    $i \leftarrow i+1;$

    *Drawn at random a codeword c from $C_w$.*

    *If $c \in L_w$ then $S \leftarrow S+1$; end if.*

*End while.*

$R(L_w) \leftarrow s/i.$

*End.*

---

To get the confidence interval of R corresponding to the precision $\mu$ we use the algorithm above at each time j to compute a value $R_j$ of R. When the process is repeated q times the value of R belongs to the interval:

$$I(R) = \left[\overset{*}{R} - \frac{\sigma.\beta}{\sqrt{q}};\ \overset{*}{R} + \frac{\sigma.\beta}{\sqrt{q}}\right] = \left[\overset{*}{R}_{left};\ \overset{*}{R}_{right}\right] \quad \text{With a probability of error equal to } \mu$$

Where:
- $\overset{*}{R} = q^{-1} \cdot \sum_{j=1}^{q} R_j$ is the average value of the recovery rate.
- $\sigma = \sqrt{\frac{1}{q-1} \sum_{j=1}^{q} (\overset{*}{R} - R_j)^2}$ is the variance of R.
- $\beta$ is the solution of the equation: $\quad \frac{2}{\sqrt{2\pi}} \int_{\beta}^{\infty} e^{-\frac{u^2}{2}} du = 1 - \mu \qquad (7)$

The table 2 gives the values of $\beta$ for some values of $\mu$ between 94% and 99.99

Table 2. Some values of β.

| μ | β |
|---|---|
| 0.9999 | 3.89 |
| 0.99 | 2.57 |
| 0.98 | 2.32 |

The confidence interval $I(|C_w|)$ of the number of code words having the weight w can be written as follows: $I(|C_w|) = \left[\frac{|L_w|}{R_{right}};\frac{|L_w|}{R_{left}}\right]$ When the variance σ is null and the recovery rate is equal to 1 we can

decide that $L_w$ is equal to $C_w$.

*3.3 Error impulse technique to find a list of codewords of a given weight.*

The automorphism group of a code C(n,k,d) is the set of permutations under which it remain invariant. For cyclic codes, this group contains all cyclic permutations of length n. When a codeword c ∈ C is available then a set of codeword having the same weight as c can be obtained by applying these permutations.

The error impulse technique (Berrou 2002, ASKALI 2012) is a fast and efficient method to find the minimum distance of linear codes. In this work we use this technique to find not only codewords of the minimum weight but a list of codewords with small weights as follow:

Let be $c^1$ a codeword to transmit over a communication channel, in the modulation phase $c^1$ is converted to the sequence $c^2$ which will transmitted and it can be perturbed by a noise e therefore the received sequence r with float symbols can be different from $c^1$. The soft decision decoding algorithms try to decode the sequence r, in general, by searching the codeword $c^2$ having the smallest Euclidean distance from r. When the decided codeword $c^2$ is not equal to $c^1$ the sum $c^3=c^1+c^2$ is a codeword having in general a small weight. When this process is iterated a list $L_w$ of code words having the weight w can be obtained.

**4. Validation of the proposed METHOD.**

*4.1 Validation of the proposed method.*

To validate the proposed method we give in the table 4 the exact partial weight enumerators and the approximated partial weight enumerators of radius 3 for some linear codes. This table shows that the proposed method yield a very good approximation (true at 100%) of a partial weight enumerators. Therefore it can be used as a powerful tool to analyze the performances of linear error correcting codes of average lengths.

Table 3. Comparison between the exact and approximated PWE.

| Code | The exact PWE | The approximated PWE |
|---|---|---|
| QR(47,24) | 11 : 4324<br>12 : 12972<br>15 : 178365 | 11 : 4324<br>12 : 12972<br>15 : 178365 |
| QR(71,36) | 11 : 497<br>12 : 2485<br>15 : 47570 | 11 : 497<br>12 : 2485<br>15 : 47570 |
| QR(73,37) | 13 : 1533<br>14 : 6570<br>15 : 19272 | 13 : 1533<br>14 : 6570<br>15 : 19272 |

When the dimension of the code C is large, the proposed method can be used to approximate the PWE of C, the table 4 gives the confidence intervals $I(|C_w|)$ regarding the approximated partial weight enumerator of the BCH(127,50,27) code generated by the polynomial:

$g_1(x) = 1 + x^1 + x^2 + x^3 + x^5 + x^6 + x^9 + x^{10} + x^{11} + x^{13} + x^{14} + x^{15} + x^{17} + x^{18} + x^{26} + x^{27} + x^{28} + x^{33} + x^{35} + x^{36} + x^{37} + x^{38} + x^{42} + x^{43} + x^{45} + x^{47} + x^{48} + x^{51} + x^{56} + x^{57} + x^{58} + x^{59} + x^{60} + x^{64} + x^{65} + x^{68} + x^{72} + x^{75} + x^{77}$.

With M=10, β=2.57, q=100. The table 5 gives a comparison between the exact PWE and its approximated value. It shows that $|C_w|$ is in general in the confidence interval or it is near to its extremities.

15

Table 4. Confidence interval $I(|C_w|)$ for the BCH (127,50) code.

| w | $|L_w|$ | $\overset{*}{R}$ | σ | $I(|C_w|)$ |
|---|---|---|---|---|
| 27 | 104 | 0,26045736 | 0.086847 | [35364; 41993] |
| 28 | 104 | 0,06536888 | 0.023571 | [140004; 168602] |
| 31 | 104 | 0,01171173 | 0.004031 | [3922283; 4683496] |
| 32 | 104 | 0,00370467 | 0.001421 | [12285596; 14972120] |

Table 5. Validation of approximated PWE for BCH (127,50).

| The exact PWE | The approximated PWE | $|C_w| \in I(|C_w|)$ |
|---|---|---|
| 27 : 40894 | 27 : 38394 | Yes |
| 28 : 146050 | 28 : 152978 | Yes |
| 31 : 4853051 | 31 : 4269224 | No |
| 32 : 14559153 | 32 : 13496466 | Yes |

*4.2 Application on some BCH codes.*

By using the proposed method with the OSD[3] decoder (Fossorier 1995) we give in the table 6 the confidence interval of a partial weight enumerators for BCH(130,66), BCH(103,47) and BCH(111,55) codes. The first code is obtained by shortening the binary primitive BCH (255,191,17) code generated by the polynomial $g_2$ and the two other codes are obtained by shortening the binary primitive BCH(127,71,19) code generated by the polynomial $g_3$.

$g_2(x) = 1 + x^2 + x^3 + x^5 + x^6 + x^9 + x^{10} + x^{11} + x^{14} + x^{15} + x^{16} + x^{22} + x^{23} + x^{24} + x^{25} + x^{26} + x^{27} + x^{31} + x^{34} + x^{35} + x^{37} + x^{39} + x^{40} + x^{42} + x^{43} + x^{45} + x^{46} + x^{47} + x^{48} + x^{49} + x^{52} + x^{53} + x^{56} + x^{58} + x^{59} + x^{60} + x^{62} + x^{63} + x^{64}$

$g_3(x) = 1 + x^4 + x^{10} + x^{11} + x^{13} + x^{16} + x^{17} + x^{20} + x^{23} + x^{24} + x^{25} + x^{28} + x^{32} + x^{35} + x^{39} + x^{40} + x^{41} + x^{42} + x^{43} + x^{44} + x^{45} + x^{46} + x^{48} + x^{49} + x^{51} + x^{53} + x^{56}$

Table 6. Confidence intervals.

| Code | Weight w | $I(|C_w|)$ |
|---|---|---|
| BCH (130,66) | 17 | [58;58] |
| | 18 | [284; 339] |
| | 19 | [1244; 14 57] |
| | 20 | [10155; 13233] |
| BCH (103,47) | 19 | [3319;24252] |
| | 20 | [14336; 27760] |
| | 21 | [17552;901580] |
| | 22 | [120862;183060] |
| BCH(111,55) | 19 | [ 20163; 27212] |
| | 20 | [78991;127621] |
| | 21 | [154269;317927] |
| | 22 | [648819;1335370] |

The confidence intervals regarding the approximated partial weight enumerator of the three BCH codes given in the table 6 are obtained by using the parameter β=3.89 all other parameters are given in the table 7 with the average values of $|C_w|$ for each weight w. By using the approximated PWE we give in the figure 5 (above) the approximated analytical performance of these codes.

Table 7. Approximated partial weight enumerator of some BCH codes.

| C | w | $|L_w|$ | R | q | σ | M | $|C_w|$ |
|---|---|---|---|---|---|---|---|
| BCH (130,66) | 17 | 58 | 1 | 30 | 0 | 10 | 58 |
| | 18 | 232 | 0.748731 | 101 | 0.169188 | 5 | 309 |
| | 19 | 911 | 0.678383 | 133 | 0.158547 | 6 | 1342 |
| | 20 | 1478 | 0.128609 | 45 | 0.029190 | 10 | 11492 |
| BCH (103,47) | 19 | 300 | 0.051368 | 3 | 0.017364 | 10 | 5840 |
| | 20 | 795 | 0.042046 | 7 | 0.009120 | 10 | 18709 |
| | 21 | 530 | 0.015391 | 3 | 0.006591 | 10 | 34434 |
| | 22 | 2351 | 0.016147 | 9 | 0.002548 | 10 | 145795 |
| BCH(111,55) | 19 | 1062 | 0.045848 | 258 | 0.028164 | 5 | 23163 |
| | 20 | 877 | 0.008987 | 76 | 0.004740 | 5 | 97583 |
| | 21 | 1002 | 0.004823 | 30 | 0.002354 | 5 | 207738 |
| | 22 | 836 | 0.000957 | 10 | 0.000269 | 5 | 873317 |

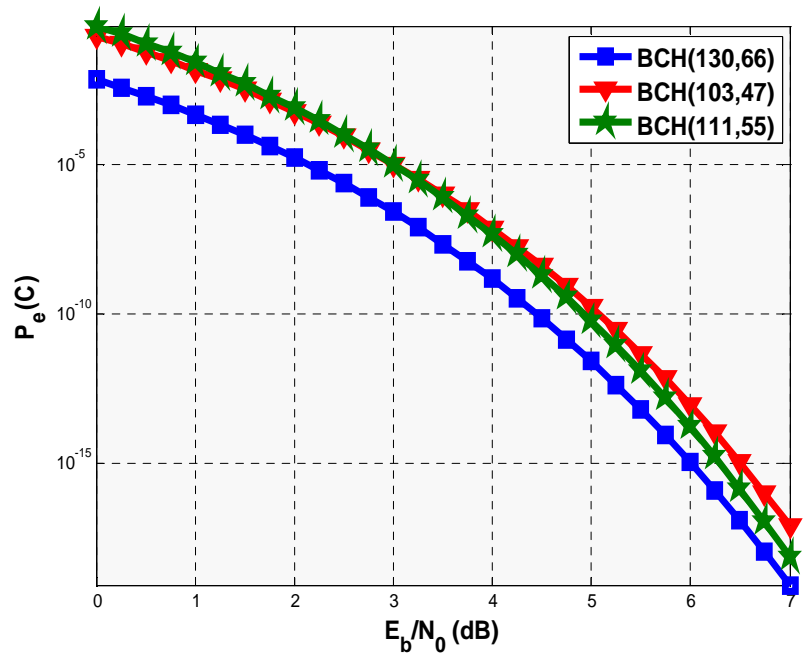

Figure 5. The analytical performances obtained by PWE for the BCH (130, 66), BCH (103, 47) and BCH (111, 55) codes.

*4.3 Run time of the proposed method.*

The table 8 shows the run time of the proposed method for two linear codes by using a simple PC configuration: AMD Athlon™ 64 processor, 1GB of RAM. It shows that we can obtain a partial weight enumerator in a few minutes for codes of average length and dimension.

Table 8. Run time of the proposed method.

| Code | Weight | Run time of the proposed method (in seconds) |
|---|---|---|
| BCH(63,39) | 9 | 73 s |
|  | 10 | 383 s |
|  | 11 | 5840 s |
| QR(71,36) | 11 | 43 s |
|  | 12 | 254 s |
|  | 15 | 10291 s |

**5. Conclusion.**

In this paper we have presented an efficient method for finding a partial weight enumerator of a linear code and therefore its analytical performance. Using the error impulse technique and some elements of the automorphism group a large list of code words of a given weight can be obtained and it can be used in a Monte Carlo method for approximating a partial weight enumerator. This method is validated on some codes with known weight enumerator and it is then used to get the partial weight enumerators and the performances of other codes with unknown weight enumerators.

**S. NOUH** – Phd student, Ecole nationale  Supérieure d'informatique et d'analyse Système, ENSIAS, Avenue Mohammed Ben Abdallah Regragui, Madinat Al Irfane, BP 713, Agdal Rabat, Maroc; e-mail: nouh_ensias@yahoo.fr. Major Fields of Scientific Research: Computer Science and Engineering. Areas of interest are Information and Coding Theory.

**B. AYLAJ** – Phd student, Université de Chouaib Doukkali, MAPI Labo, Faculté des sciences, Route Ben Maachou, 24000, El jadida, Maroc; e-mail: bouchaib_aylaj@yahoo.fr. Major Fields of Scientific Research: Computer Science and Engineering. Areas of interest are Information and Coding Theory.

**M. Belkasmi** – Professor, Ecole nationale  Supérieure d'informatique et d'analyse Système, ENSIAS, Avenue Mohammed Ben Abdallah Regragui, Madinat Al Irfane, BP 713, Agdal Rabat, Maroc; e-mail: belkasmi@ensias.ma. Major Fields of Scientific Research: Computer Science and Engineering. Areas of interest are Information and Coding Theory.